\documentclass[a4paper,12pt]{article}

\usepackage{graphicx} 
\usepackage{amsmath} 
\usepackage{hyperref} 
\usepackage{color} 
\usepackage{bm} 
\usepackage{lipsum} 
\usepackage{float} 
\usepackage{booktabs} 
\usepackage{siunitx} 
\usepackage{amsfonts} 
\usepackage{amssymb} 
\usepackage{geometry} 
\usepackage{fancyhdr} 

\title{Model of an Open, Decentralized Computational Network with Incentive-Based Load Balancing\thanks{A conceptual framework for discussion and proof of concept, exploring its applicability.}}
\author{German Rodikov\thanks{GR acknowledges financial support from the Italian Ministry MUR under the PRIN2020 project "Dynamic models for a fast-changing world: An observation driven approach to time-varying parameters" (grant agreement no. 20205J2WZ4).} \\
University of Bologna}
\date{1 January 2025 }

\begin{document}

\maketitle
\begin{abstract}
This paper proposes a model that enables permissionless and decentralized networks for complex computations. We explore the integration and optimize load balancing in an open, decentralized computational network. Our model leverages economic incentives and reputation-based mechanisms to dynamically allocate tasks between operators and coprocessors. This approach eliminates the need for specialized hardware or software, thereby reducing operational costs and complexities. We present a mathematical model that enhances restaking processes in blockchain systems by enabling operators to delegate complex tasks to coprocessors. The model's effectiveness is demonstrated through experimental simulations, showcasing its ability to optimize reward distribution, enhance security, and improve operational efficiency. 

Our approach facilitates a more flexible and scalable network through the use of economic commitments, adaptable dynamic rating models, and a coprocessor load incentivization system. Supported by experimental simulations, the model demonstrates its capability to optimize resource allocation, enhance system resilience, and reduce operational risks. This ensures significant improvements in both security and cost-efficiency for the blockchain ecosystem.
    
\end{abstract}

\section{Introduction}
Blockchain technologies have continually evolved, focusing on enhancing scalability and security. Innovations such as the Optimistic Delegation Framework (ODF) and Proposer-Builder Separation (PBS) refine transaction validation processes by assigning specialized roles within blockchain networks. This paper presents an optimization model that aligns incentives and distributes computational tasks effectively among Operators and Coprocessors within the AVS, leveraging a restaking mechanism. It explores how restaking operators use standalone coprocessors for complex tasks, reducing the need for specialized hardware or software and cutting securitization costs while boosting network efficiency.

The main assumptions and model considerations include: a reward function that compensates Operators and Coprocessors based on task complexity and computational resources; slashing conditions that define penalties for non-compliance; a collateral system to secure commitments; a reputation system that affects task allocation and reward distribution as the effect of it; an auction mechanism for cost-effective Coprocessor selection; and a risk analysis framework to manage potential losses effectively.

Our approach uses probabilistic modeling to predict and manage the risks associated with task execution failures, forming the basis for our optimization problem aimed at maximizing total expected rewards while mitigating slashing risks. The model's dynamic nature allows it to adapt over time, capturing the temporal changes in network conditions and participant behavior through a feedback mechanism that updates probabilities based on past performance data.

To ensure the model's practical relevance and adaptability, we incorporate mechanisms for demand forecasting and capacity planning, essential for managing resource allocation in response to fluctuating demands for AVS tasks. This paper outlines the development of these models, providing a framework for their integration into existing blockchain infrastructures to enhance efficiency and economic viability.

\subsection{Preliminaries}
In this section, we provide a deeper understanding of the existing system's operation and the foundational changes proposed by our model. This study develops a comprehensive mathematical model for the Optimistic Delegation Framework (ODF) and Proposer-Builder Separation (PBS) with a focus on Actively Validated Services (AVS) within the blockchain ecosystem. Our model seeks to optimize the distribution of rewards and minimize risks associated with penalizations, known as "slashing," under these frameworks. 

In the current PBS model, the roles of proposers and builders are separated to reduce potential conflicts of interest and enhance transaction censorship resistance. Proposers are responsible for adding new blocks to the blockchain, while builders, operating independently, focus on constructing the content of these blocks, including transaction ordering. This separation ensures that builders can specialize in maximizing block value through sophisticated strategies without directly influencing block proposals.

The model introduces a layer of complexity and opportunity by enabling operators to delegate specific computational tasks to specialized coprocessors. This delegation is critical for handling tasks that are resource-intensive or require specialized computational capabilities, which are beyond the efficient handling capacity of general operators.

\vspace{1cm}
\textbf{System Architecture and Role Descriptions:}
\begin{itemize}
    \item \textbf{Operators}  perform validations and execute tasks as part of the AVS, ensuring adherence to distributed validation semantics. These tasks can range from lightweight to heavyweight, depending on the trust model and operational demands of the specific AVS they opt into. Operators plays a pivotal role by ensuring the integrity of transactions and deciding whether to execute tasks directly or delegate them to coprocessors. Their operations are augmented by several methods including reward calculation, coprocessor selection based on a scoring system that considers computational capability and reliability, and performance updates that influence future task allocations and reward calculations.
    \item \textbf{Coprocessors} are specialized nodes that handle delegated tasks. They participate in auctions to win tasks by making competitive bids that reflect their processing capabilities and current workload. The efficiency of coprocessors directly impacts their profitability and the overall efficiency of the blockchain operation.
    \item \textbf{AVS Tasks} within this system vary in complexity and associated risks. The model takes into account the difficulty of tasks, potential rewards, and risks of penalties to optimize task allocation.
\end{itemize}

In this paper, we investigate the use of standalone coprocessors by restaking operators for complex computations. This model enables operators to effectively execute AVS tasks without the need to personally manage specific hardware or software, reducing the costs associated with securitization and enhancing operational efficiency within the network.

Our model implements a dynamic and adaptive mechanism to adjust task allocations based on real-time performance data. This approach not only aims to balance the computational load across the network but also to refine the reward system based on evolving network conditions and coprocessor performance. Additionally, the proposed model incorporates a feedback mechanism to continuously learn from past allocations and outcomes, thus improving its predictive capabilities and efficiency.

Dynamic Task Allocation, by integrating real-time performance analytics, the system can dynamically allocate tasks to operators and coprocessors based on their current load and historical accuracy.
Feedback Loop, implementing a feedback loop will allow the system to adapt its parameters based on the outcomes of previous task allocations, enhancing both system resilience and task performance predictability.
Auction Mechanisms, the auction process for selecting coprocessors will be refined to include factors such as past performance, reputation, and economic bidding strategies to ensure fair and efficient task distribution.

These enhancements are designed to leverage the specialized capabilities of coprocessors more effectively while maintaining the decentralized nature and security of the blockchain.

\section{Methodology}
The proposed model is formulated as a stochastic optimization problem where the objective is to maximize the expected net rewards of Operators and Coprocessors, taking into account the risk of penalties and the constraints of the system.

\subsection{Variables and Parameters}
The model comprises the following entities and their respective variables:
\begin{itemize}
    \item $V = \{v_1, v_2, \ldots, v_n\}$: the set of all Operators.
    \item $C = \{c_1, c_2, \ldots, c_m\}$: the set of all Coprocessors.
    \item $A = \{a_1, a_2, \ldots, a_k\}$: the set of all tasks within the AVS.
    \item $r_{v,a}$: the reward for Operator $v$ upon successful completion of task $a$.
    \item $s_{v,a}$: the slashing risk for Operator $v$ when task $a$ is performed incorrectly.
    \item $p_{v,c,a}$: the probability of successful task completion by Operator $v$ and Coprocessor $c$ for task $a$.
    \item $b_{c,a}$: the bid submitted by Coprocessor $c$ for task $a$.
    \item $l_{c,a}$: the collateral posted by Coprocessor $c$ for task $a$.
\end{itemize}

\subsection{Objective Function}
The objective function is designed to maximize the total expected benefit for each Operator, computed as follows:
\[
B_v = \sum_{a \in A} \left( r_{v,a} \cdot p_{v,c,a} - s_{v,a} \cdot (1-p_{v,c,a}) \right)
\]
where $B_v$ represents the net expected benefit for Operator $v$, accounting for both rewards and risks associated with task executions.

\subsection{Optimization Problem}
The selection of Coprocessors by Operators is modeled through an auction mechanism, incorporating both cost considerations and the probability of successful task completion. The optimization problem can be stated as:
\[
\text{Maximize} \quad \sum_{c \in C} \sum_{a \in A} x_{v, c, a} \left( r_{v, a} \cdot p_{v, c, a} - s_{v, a} \cdot (1-p_{v, c, a}) - b_{c, a} \right)
\]
subject to:
\begin{align*}
\sum_{c \in C} x_{v, c, a} & \leq 1 \quad \forall a \in A \\
x_{v, c, a} \cdot l_{c, a} & \geq \text{Minimum Collateral Requirement} \quad \forall c \in C, a \in A \\
x_{v, c, a} & \in \{0,1\} \quad \forall v \in V, c \in C, a \in A
\end{align*}
Here, $x_{v,c,a}$ is a binary variable indicating whether Operator $v$ selects Coprocessor $c$ for task $a$.

\begin{figure}[H]
    \centering
    \includegraphics[width=1.0\textwidth]{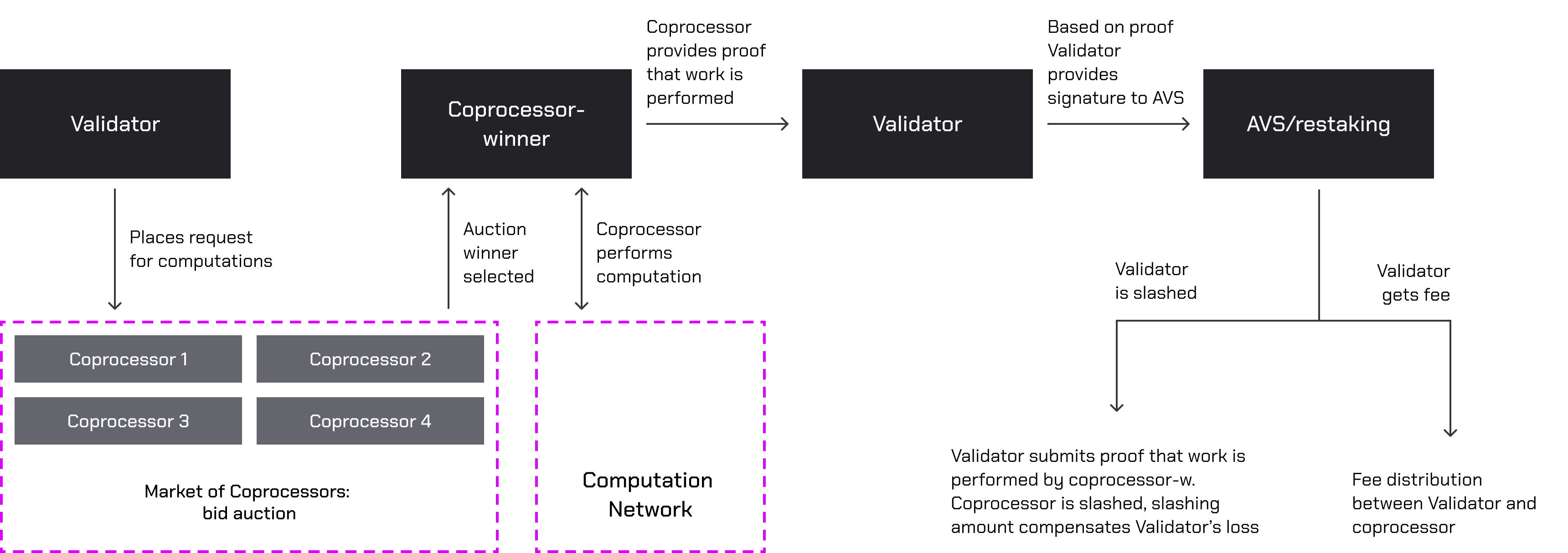}
    \caption{Precompute auction model based on bids}
    \label{fig:schema}
\end{figure}

In addition to the aforementioned stochastic optimization framework, our study further explores the application of economic incentive structures within blockchain ecosystems. To demonstrate the practical implications and robustness of our model, we simulate the Economic Incentive Optimal Load Curve \ref{sec:load_curve}. Through detailed simulations, we illustrate the effectiveness of the model in maintaining an optimal balance between resource allocation and economic incentives, which is crucial for sustaining high network performance and reliability.

To accommodate the evolving nature of the blockchain ecosystem, the model integrates dynamic elements through time-dependent variables, allowing it to adapt and respond to changing conditions over time. This dynamic approach ensures that strategies can be adjusted based on past performance and anticipated future conditions, thus enhancing the robustness and effectiveness of the model.

\section{Experiment}

\subsection{Experiment Setup}

In our experimental setup, we simulate a blockchain network consisting of operators and coprocessors to test the proposed optimization model described earlier. The simulation aims to understand how the system behaves under various scenarios, particularly focusing on the distribution of tasks, the impact of reward systems, and the effectiveness of setup mechanisms.

\subsubsection{Configuration Parameters}
\begin{itemize}
    \item \textbf{Operators}: 100
    \item \textbf{Coprocessors}: 100
    \item \textbf{Simulation Periods}: 1000
\end{itemize}

Each operator and coprocessor is initialized with a set of attributes. Operators have varying resources, reputations, and preferences on a task. We have over 1 million tasks to compute across all periods. Coprocessors are differentiated by their resources and reputations. Tasks vary in difficulty, potential rewards, associated slashing penalties, and risk factors.

\subsection{Methodology}

\subsubsection{Task Allocation and Execution}
Operators decide on task assignments based on their resources and the complexity of the tasks. They may choose to execute tasks themselves or auction them to coprocessors depending on their strategy, which considers both task difficulty and potential rewards.

\begin{enumerate}
    \item \textbf{Task Execution by Operators}:
    Operators execute tasks if their available resources exceed the task difficulty and the task complexity.
  
    \item \textbf{Auctioning to Coprocessors}:
    For more complex tasks, operators auction these tasks to coprocessors. The auction mechanism is based on bids reflecting the coprocessors' capabilities and current workload.
\end{enumerate}

\subsubsection{Auctioning to Coprocessors}
For tasks that exceed the complexity threshold of operators, an auction mechanism is employed to determine which coprocessor will undertake the task. This auction is a modified Dutch auction, where the initial high bid price is progressively lowered until a coprocessor accepts the task. However, unlike traditional Dutch auctions that focus solely on price, this system also integrates the coprocessors' resources and current workload into the bidding process.

\paragraph{Dutch Auction Mechanics}
In the Dutch auction model used in our system:
\begin{itemize}
    \item The auction starts with a high price set close to the maximum reward associated with the task.
    \item The price is progressively lowered at fixed intervals until a coprocessor places a bid, signaling their willingness to accept the task at the current price.
\end{itemize}

\paragraph{Integration of Coprocessor Resources}
The unique aspect of our auction model is the integration of each coprocessor’s available resources and workload:
\begin{itemize}
    \item \textbf{Resource Consideration}: Each coprocessor's current available resources are considered when setting the initial and decrementing prices. Coprocessors with higher available resources may qualify for starting the bidding at more favorable prices, as they are more capable of efficiently completing the task.
    \item \textbf{Workload Consideration}: The current workload of a coprocessor affects their ability to bid. Coprocessors nearing or at their capacity might be disincentivized to participate early in the auction, or they might bid at lower prices that reflect their higher operational risks and costs.
\end{itemize}

\paragraph{Bidding Strategy}
The strategy for each coprocessor is to balance their resource utilization with the potential rewards:
\begin{itemize}
    \item Coprocessors must decide whether to bid early at a higher price, risking a decrease in available resources but gaining higher rewards.
    \item Alternatively, they might wait for the price to drop to lower levels that better match their current resource availability and risk profile.
\end{itemize}

This modified Dutch auction system ensures that tasks are allocated not only based on price competitiveness but also in a manner that respects the operational capabilities and current conditions of each coprocessor. This strategic allocation aids in maintaining system efficiency and task reliability, even under varying network conditions.

\subsection{Analysis Approach}

The collected data will be analyzed to assess:
\begin{enumerate}
    \item \textbf{Efficiency of Task Distribution}:
    How effectively tasks are allocated between operators and coprocessors based on their resources and capabilities.
  
    \item \textbf{Economic Incentives}:
    The impact of reward structures and slashing mechanisms on the behavior of operators and coprocessors.
  
    \item \textbf{System Robustness}:
    How the system withstands various operational stresses, such as high task complexity and resource constraints.

    \item \textbf{Dynamic Adaptation}:
    The responsiveness of the system to changing conditions through the feedback mechanism, which adjusts operational strategies based on past outcomes.
\end{enumerate}

This experiment is designed to validate the hypothesis that the proposed model enhances the efficiency and security of blockchain operations through optimized Proposer-Builder Separation:
\begin{itemize}
    \item A balanced distribution of tasks that maximizes the utilization of available resources while minimizing risks and penalties.
    \item Adaptive behaviors in operators and coprocessors, leading to an evolving and resilient blockchain ecosystem.
    \item  Investigating the stability of the system to determine optimal parameters ensuring its sustained existence.
\end{itemize}

By analyzing the simulation results, we aim to demonstrate the practical viability of our model and its potential to improve blockchain operations.

\section{Results}
This section presents the empirical outcomes of the experimental simulations performed to validate the efficacy of the proposed optimization model within the blockchain's restaking mechanism. The results not only illustrate the impact of various slashing factors on the system's stability and reward mechanisms but also reveal critical insights into the operational dynamics between operators and coprocessors under different network conditions. By systematically analyzing these outcomes, we can derive significant conclusions about the robustness of the proposed model and its adaptability to the fluctuating demands of a blockchain environment.

\subsection{Impact of Slashing Factors on Operators Dynamics and Reward Distribution}

The experiment investigates the effects of slashing factors on the activity of operators and their reward accumulation over time. Slashing, a punitive measure, is implemented to deter malicious activities by penalizing operators and coprocessors for misbehavior or failures. This section analyzes how different slashing intensities, represented by the factor $s$, influence the network's health and economic incentives.  This section specifically focuses on identifying some non-zero slashing parameter, observing the long-term survival and reputation of coprocessors, and showing the overall stability of the system.

\begin{figure}[H]
    \centering
     \includegraphics[width=0.8\textwidth]{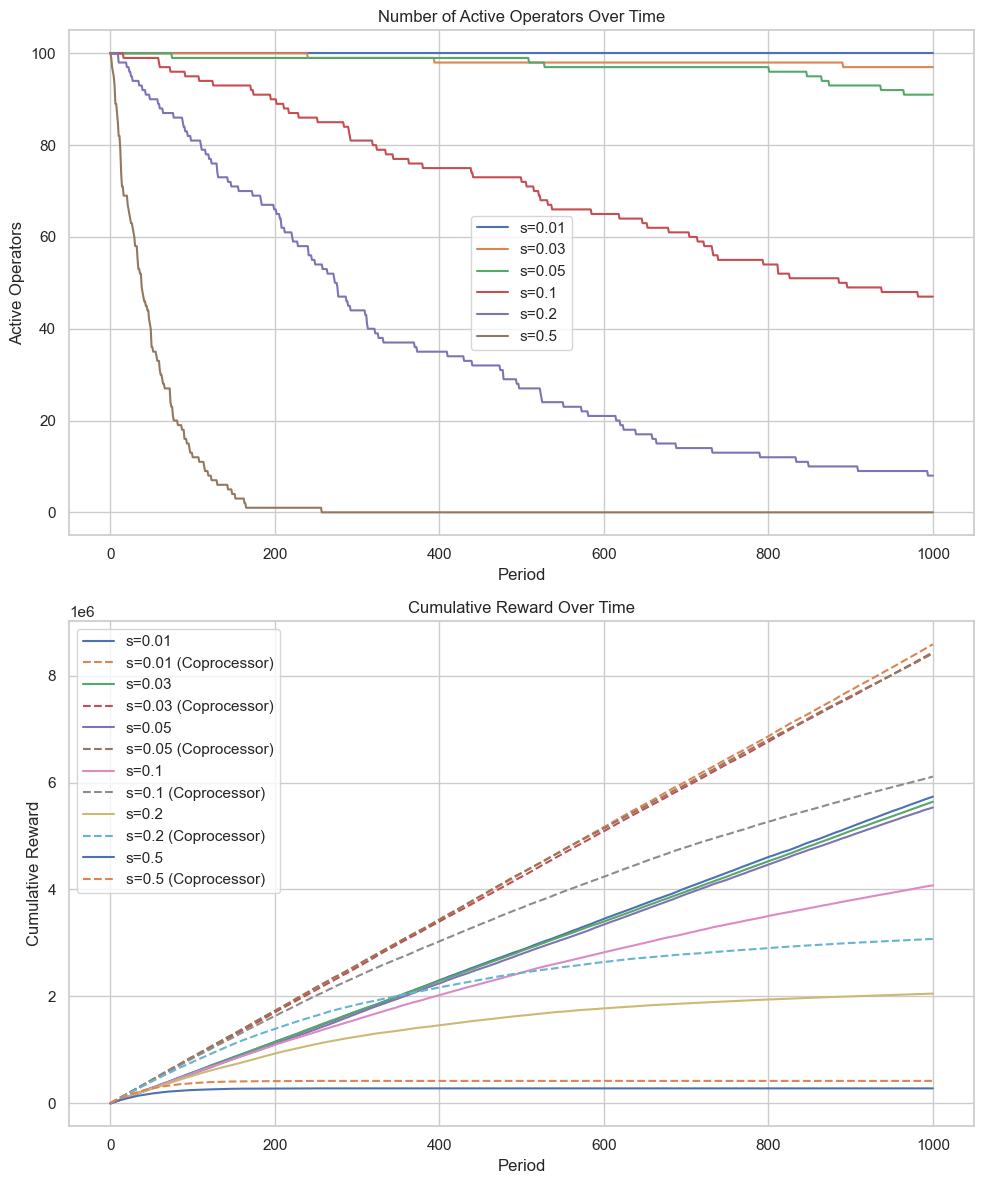}
    \caption{Top: Number of Active Operators Over Time; Bottom: Cumulative Reward Over Time, segmented by different slashing factors $s$.}
    \label{fig:slashing_impact}
\end{figure}

\subsubsection{Number of Active Operators}
The top plot in Figure~\ref{fig:slashing_impact} shows the number of active operators over time under different slashing factors. Notably, higher slashing factors ($s = 0.2, 0.5$) lead to a rapid decline in activity in the system, indicating a harsh punitive environment that may drive operators away from the network. Conversely, very low slashing factors ($s = 0.01$) maintain a higher number of active operators, suggesting insufficient deterrence against potential misbehavior. 

\subsubsection{Cumulative Reward Over Time}
The bottom plot illustrates the cumulative rewards accrued by operators and coprocessors over time, differentiated by slashing factors. It can be observed that moderate slashing factors ($\alpha = 0.03, 0.1$) strike a balance, leading to a steady growth in cumulative rewards without significant drops in operator activity. This suggests that finding an optimal slashing factor is crucial to maintaining a healthy balance between punitive measures and incentives, ensuring network security and operators/coprocessors profitability.

Our study acknowledges the limitation of simulating a fixed number of operators and coprocessors, which might not fully represent real-world dynamics where these numbers can fluctuate. In future work, we aim to extend our model to include variable operator and coprocessor counts, enhancing the robustness and applicability of our findings. This adjustment will allow us to explore more deeply how network dynamics affect blockchain efficiency and security under different regulatory frameworks.

\subsection{Analysis of Rewards Distribution Over Time}
\begin{figure}[H]
    \centering
     \includegraphics[width=0.8\textwidth]{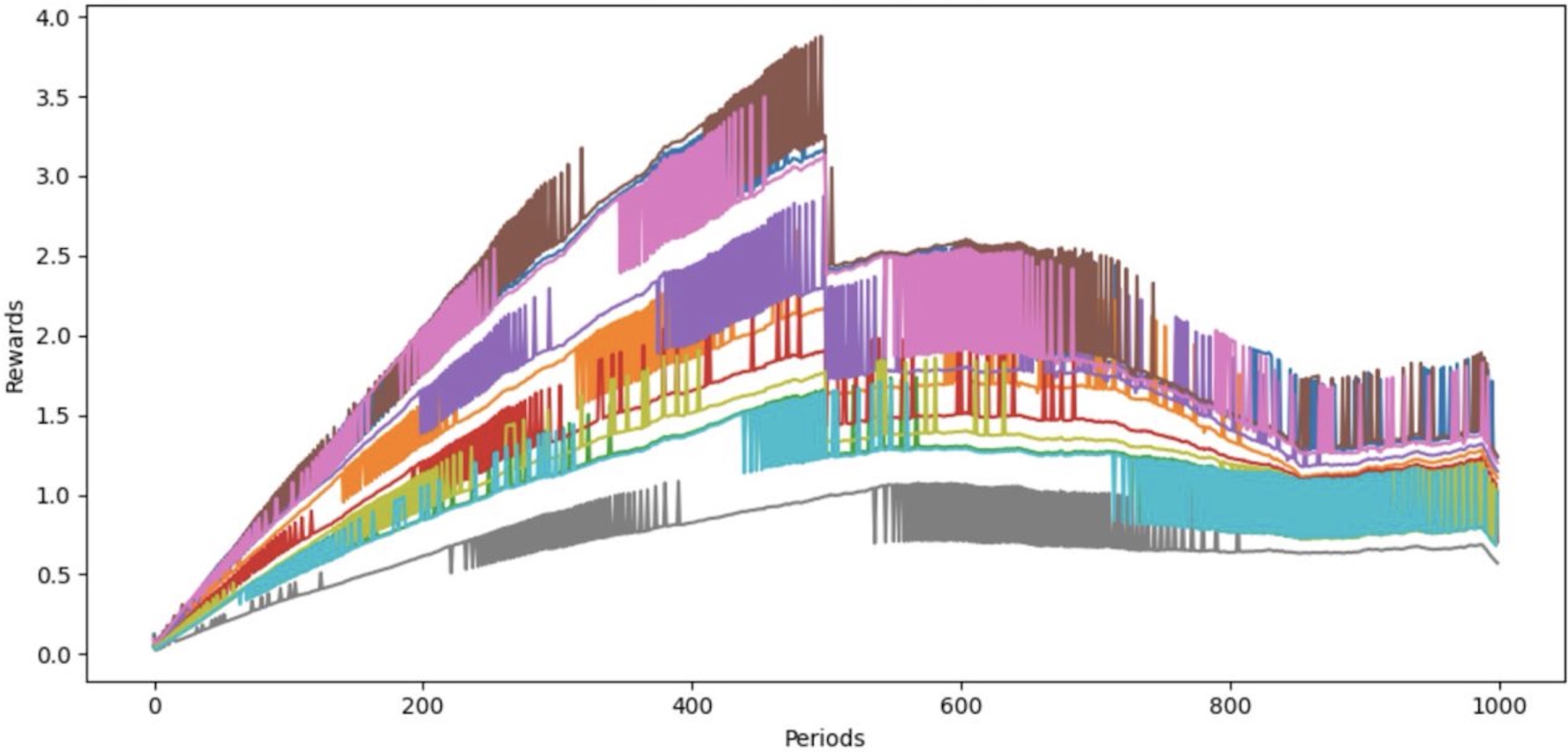}
    \caption{Subset of Operators Reward Over Time.}
    \label{fig:reward_over_time}
\end{figure}

The plot in Figure~\ref{fig:reward_over_time} provided a graphical representation that illustrates the evolution of reward distribution among operators across different periods within the blockchain environment. Initially, there is a noticeable diversity in rewards, which likely reflects a varied distribution of task complexities handled by the operators themselves. As the periods progress, the rewards across operators begin to converge towards a more uniform distribution. This trend suggests a systematic shift in the task management strategy, where operators increasingly delegate the more computationally intensive tasks to coprocessors.

Such a shift is indicative of the operators' adaptation to optimize their operational efficiency and resource utilization, leveraging the specialized capabilities of coprocessors to handle complex calculations.  Over time, as operators delegate heavier computational tasks to coprocessors, they can focus on less resource-intensive tasks. This delegation enhances the overall system stability by reducing the burden on individual operators, thereby minimizing the risk of bottlenecks or failures due to overload.

\subsection{Operator Reputation vs. Total Reward}
\begin{figure}[H]
    \centering
    \includegraphics[width=\linewidth]{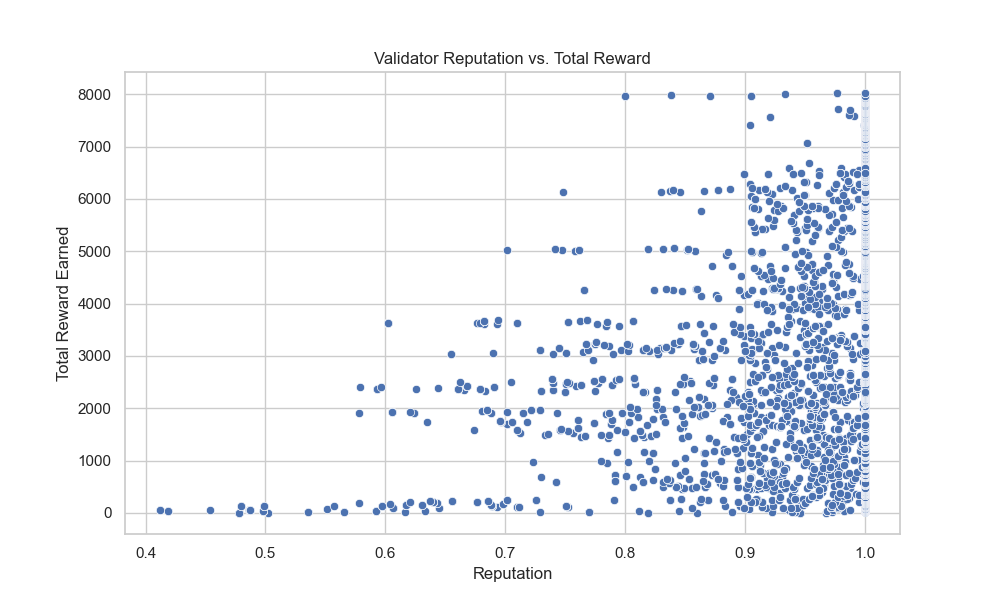}
    \caption{Scatter plot showing the correlation between operator reputation and the total rewards earned.}
    \label{fig:reputation_vs_reward}
\end{figure}

The scatter plot in Figure~\ref{fig:reputation_vs_reward} illustrates the relationship between the reputation of operators and the total rewards they earn. It is evident that operators with higher reputations tend to accumulate higher rewards, indicating that reputation is a significant factor in determining the allocation and successful completion of tasks.

\subsection{Network of Operators and Coprocessors}
\begin{figure}[H]
    \centering
    \includegraphics[width=\linewidth]{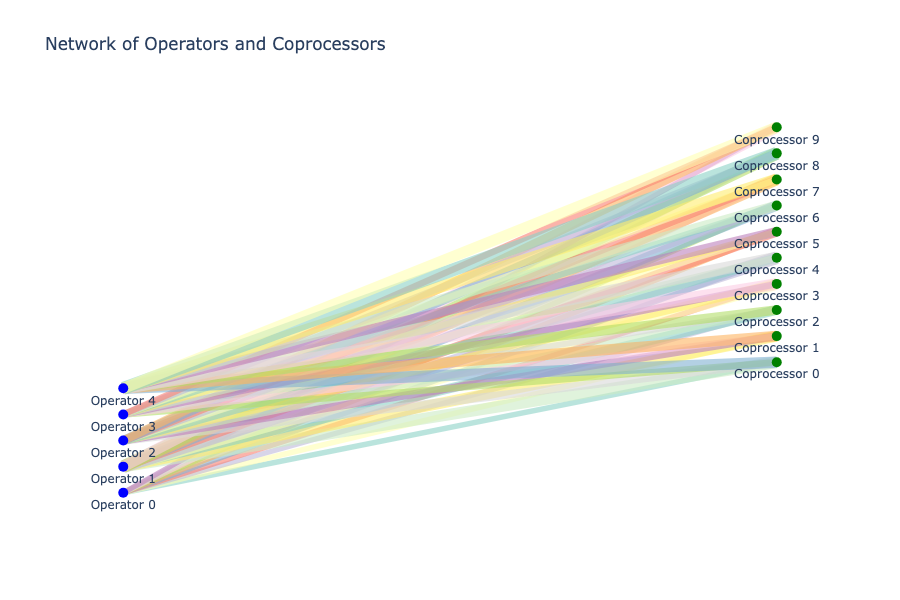}
    \caption{Network visualization of a subset of five operators and ten coprocessors illustrating the distribution of tasks.}
    \label{fig:network}
\end{figure}

Figure~\ref{fig:network} depicts the interactions between operators and coprocessors. Each line represents tasks assigned from operators to coprocessors, with different colors indicating the volume of transactions. This network diagram is crucial for understanding the flow of tasks and the dynamics of task allocation across the system.

\subsection{Coprocessor Load vs. Reward}
\begin{figure}[H]
    \centering
    \includegraphics[width=\linewidth]{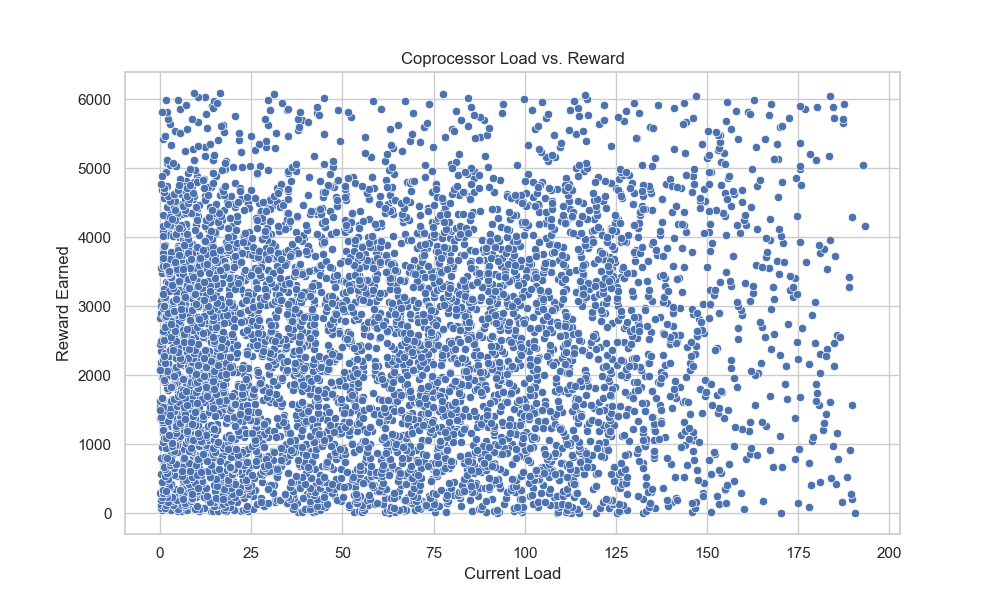}
    \caption{Plot of coprocessor load versus the rewards earned, highlighting the impact of workload on profitability.}
    \label{fig:load_vs_reward}
\end{figure}

The distribution of rewards relative to the workload of coprocessors is shown in Figure~\ref{fig:load_vs_reward}. The plot suggests that while the reward mechanism is generally effective, there is considerable variance in rewards, particularly at higher workloads, reflecting the challenges coprocessors face in balancing efficiency and reward optimization.

\subsection{Distribution of Tasks Among Coprocessors Over Time}
\begin{figure}[H]
    \centering
    \includegraphics[width=\linewidth]{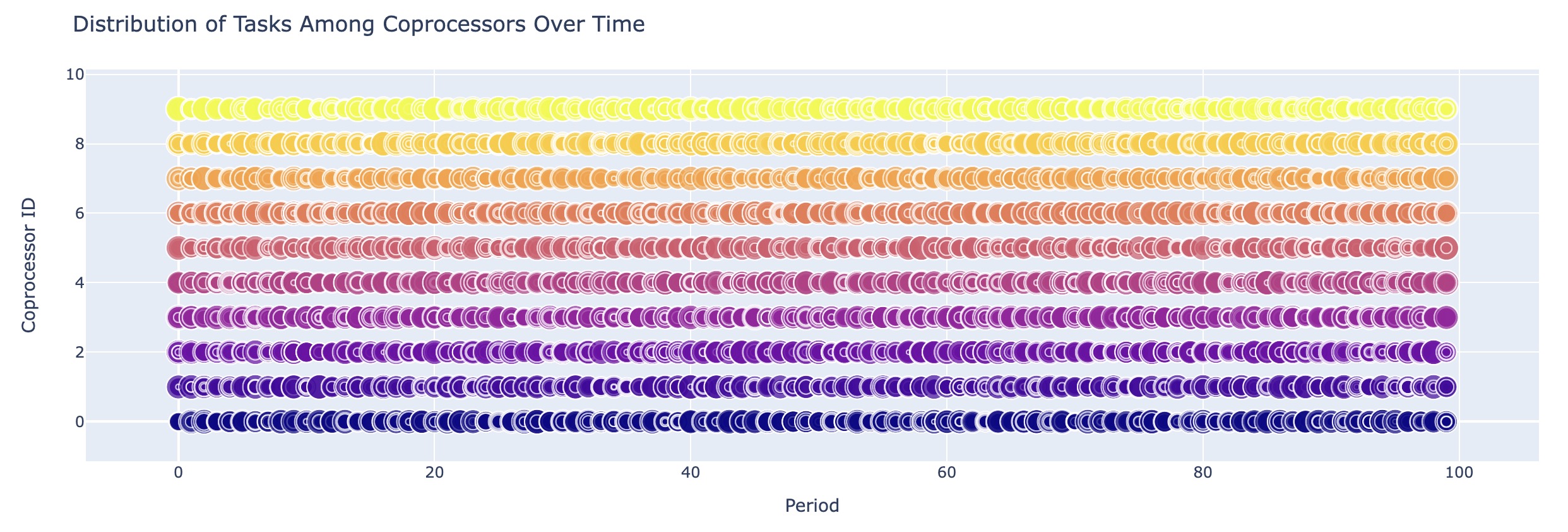}
    \caption{Color-coded visualization of tasks assigned to a subset of coprocessors over time, showcasing the operational dynamics of the network.}
    \label{fig:distribution}
\end{figure}

The dynamic allocation of tasks to coprocessors over time is visualized in Figure~\ref{fig:distribution}. This representation helps in understanding how different coprocessors are utilized throughout the operational period, emphasizing the scalability and flexibility of the network.

\clearpage
\newpage
\section{Simulation of Economic Incentive Optimal Load Curve}  \label{sec:load_curve}
The proposed approach utilizes an economic incentive optimal load curve as a permissionless protocol through a Generalized Autoregressive Score (GAS) reputation system with economic incentives. This system dynamically balances load distribution among operators and coprocessors to optimize resource utilization and enhance overall network efficiency.

\subsection{Economic Incentives and Load Management}
The GAS model is employed to create a reputation system that incentivizes coprocessors based on their performance and load management. The approach encourages low-load scenarios with positive incentives such as economic rewards for maintaining unloaded uptime. Conversely, high-load scenarios are disincentivized through amplified slashing, ensuring that coprocessors manage their resources efficiently to avoid penalties.

The optimal load curve is mathematically defined by the piecewise function:
$$
f(x) = \begin{cases}
A \cdot e^{-\left(\frac{x-B}{B}\right)^2}-C, & \text{if } x < B \\
D+(x-B) \cdot E, & \text{if } B \leq x \leq F \\
\max \left(G \cdot(x-H)^2+I, 0\right), & \text{if } x > F
\end{cases}
$$
where:
- $A, B, C, D, E, F, G, H$, and $I$ are constants derived from network parameters.
- $x$ represents the load.

\subsection{Incorporation of GAS Models}
The GAS model's core principle is the use of a scaled score to drive time variation in parameters of incentive. This is represented by the recursion:
$$
f_{t+1} = \omega + \beta f_t + \alpha S\left(f_t\right)\left[\frac{\partial \log p\left(y_t \mid f_t\right)}{\partial f_t}\right]
$$
where:
- $f_t$ is a time-varying parameter linked to load.
- $p\left(y_t \mid f_t\right)$ is the conditional observation density for observations $y_t$.
- $S\left(f_t\right)$ is a scaling function for the score of the log observation density.

This approach links the conditional observation density directly to the dynamics of $f_t$, allowing the model to adapt to large values in the data distribution without overreacting to outliers.

\begin{figure}[ht]
    \centering
    \includegraphics[width=\linewidth]{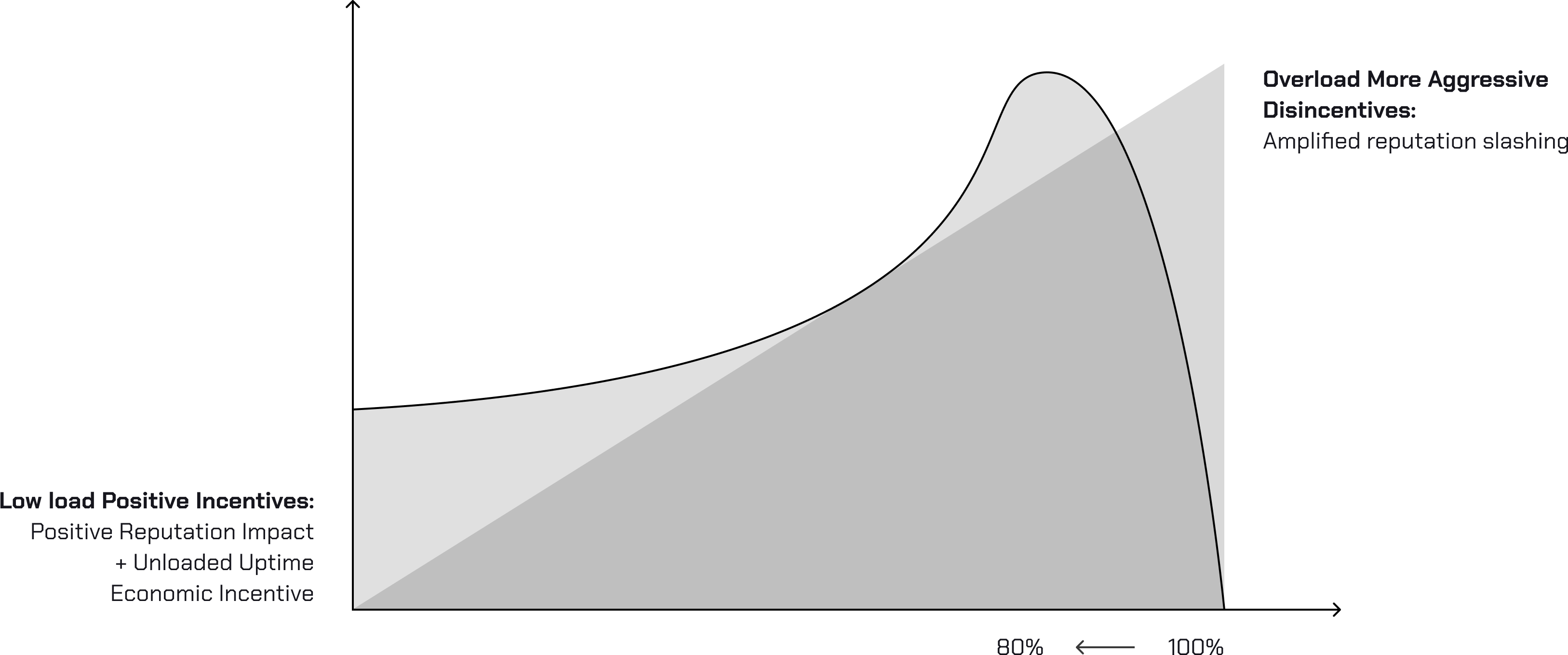}
    \caption{Theoretical Coprocessor Load}
    \label{fig:theor_Coprocessor_Load}
\end{figure}

The network diagram in Figure~\ref{fig:theor_Coprocessor_Load} illustrates the theoretical inclination of coprocessor load, highlighting the distribution of tasks and the overall network dynamics. The efficient task allocation ensures that the network remains resilient and capable of handling varying loads and operational demands.

In the simulation, coprocessors decide on auction and task bidding and execution based on their current load and reputation levels. Tasks that are less resource-intensive are executed by operators, while more complex tasks are auctioned to coprocessors. This dynamic task allocation mechanism ensures optimal resource utilization and maximizes overall network performance.

\begin{figure}[ht]
    \centering
    \includegraphics[width=\linewidth]{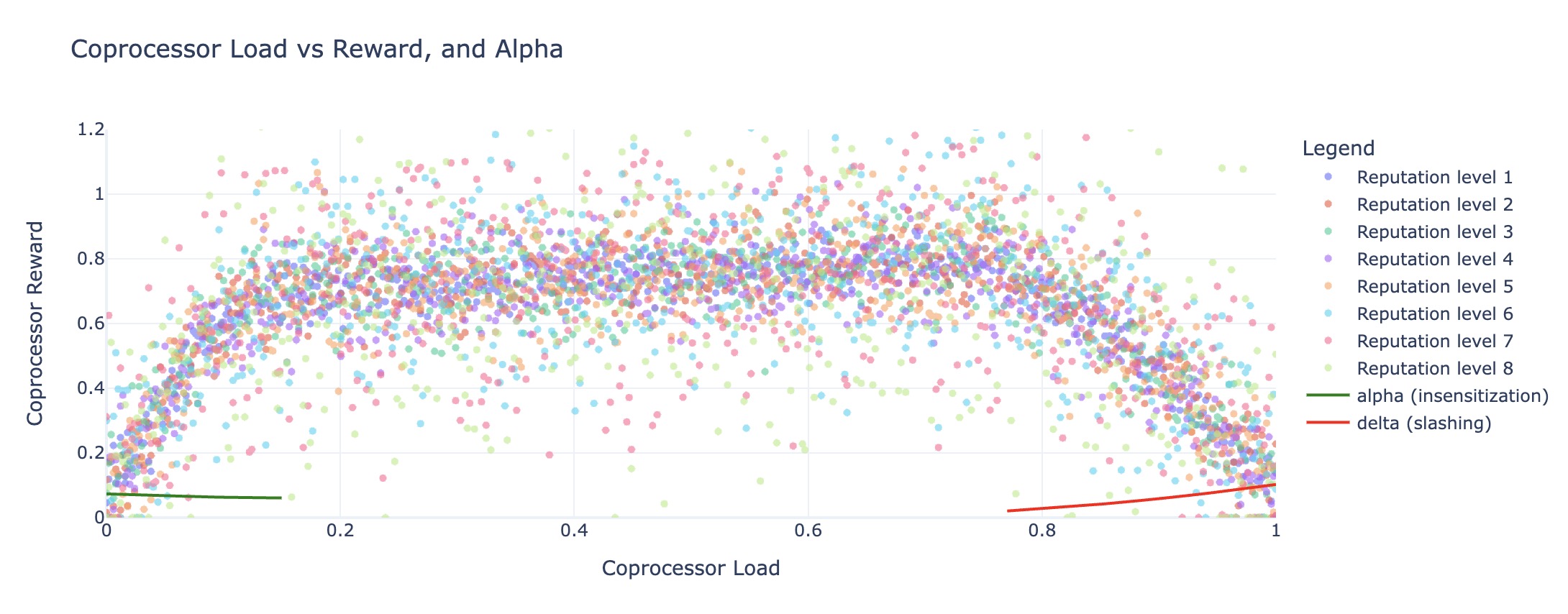}
    \caption{Simulated Coprocessor Load}
    \label{fig:simulate_Coprocessor_Load}
\end{figure}

The effectiveness of this model is demonstrated through a simulation of coprocessor load versus reward and alpha, as depicted in the scatter plot in Figure~\ref{fig:simulate_Coprocessor_Load}. The plot showcases how the load impacts the rewards earned by coprocessors and highlights the critical balance between load management and economic incentives.

This approach in the blockchain topology provides a robust framework for optimizing load distribution and economic incentives. By dynamically adjusting task allocations and incorporating reputation-based rewards and penalties, the system ensures efficient resource utilization and enhances overall network stability. The simulation results validate the model's efficacy, demonstrating its potential to improve blockchain operations, resilience, and ecosystem efficiency.

\section{Open Questions}

\subsection{Adaptive Mechanisms and Feedback Loop}
The model includes adaptive mechanisms such as real-time performance adjustments and strategic task auctions to manage resource allocation and mitigate risks. Validators and coprocessors operate within a feedback loop structure that adjusts future task allocations based on historical performance data and the generalized autoregressive score framework, ensuring a resilient blockchain ecosystem. The feedback loop ensures efficient task distribution among all participants and plays a crucial role in decision-making for delegating specific tasks to particular coprocessors.

However, there is potential for further improvement by implementing a more complex system for the feedback loop. Incorporating additional elements such as staking mechanisms can enhance the system's robustness and efficiency. 
\textbf{Open Questions:}
\begin{itemize}
\item What parameters should be considered to fine-tune the slashing mechanism to deter misbehavior while maintaining active participation?
\item How can the reward distribution model be optimized to ensure validators and coprocessors are adequately motivated?
\end{itemize}

\subsection{Error Detection and Tolerance}
The error tolerance system is designed to allow for determinism in task execution, managing errors without compromising the network's integrity and efficiency. The model employs a probabilistic approach to predict and manage risks associated with task execution failures, forming the basis for optimizing rewards and minimizing slashing risks. Ensuring accuracy and reliability in blockchain computations is crucial for maintaining network integrity. One of the challenges is how to convert a non-arithmetic answer into a deterministic result for calculations (Sanka et al. 2021).

\textbf{Open Questions:}
\begin{itemize}
\item How can independent validators, such as fishermen, or multiple coprocessors performing the same computations enhance validation efficiency?  The fishermen act as independent caretakers who monitor the network for malicious activities or errors, ensuring the integrity of the overall system.

\textbf{Fishermen Mechanism:} Independent validators, can be incorporated to monitor task execution. These validators can identify discrepancies or malicious behavior, providing an additional layer of security  (Burdges et al. 2020).

\item How can we verify that attributes such as "blue" and "tall" are consistent for items like "sofa"? Implementing an error tolerance system that allows minor variances within set limits can address this, acknowledging non-determinism while maintaining integrity. Employing diverse validation mechanisms, such as using tokens, ensures comprehensive verification.
\end{itemize}

One of the innovative technologies that can enhance validation efficiency and ensure privacy is the use of Zero-Knowledge Succinct Non-Interactive Arguments of Knowledge (zk-SNARKs). These cryptographic proofs allow one party (the prover) to prove to another party (the verifier) that they know a value or that a certain statement is true, without revealing any additional information. Employing zk-SNARKs can reduce the communication overhead and increase the speed of validation processes, significantly improving the efficiency of the system while maintaining high standards of privacy and security.

\clearpage
\newpage
\section{Conclusion}

In conclusion, this study developed and tested a mathematical model designed specifically for enhancing restaking processes within blockchain systems. Our model focuses on empowering restaking operators to utilize  coprocessors for handling complex computations, without requiring direct management of specialized hardware or software. This approach allows operators to efficiently engage in restaking by outsourcing compute-heavy AVS tasks to partner network services, thus avoiding the additional costs and complexities associated with securitization.

Our experimental simulations provide robust support for the model, demonstrating its ability to optimize reward distribution and minimize the risks associated with slashing. This is achieved through dynamic and strategic allocation of computational tasks between operators and coprocessors, which adapts seamlessly to the changing demands of the blockchain environment. The results indicate not only an increase in net expected rewards for participants but also a fair and efficient distribution of tasks, validating the efficacy of our approach.

Moreover, the model incorporates adaptive mechanisms such as real-time performance adjustments and strategic task auctions, which have proven effective in managing resource allocation and mitigating risks. These features ensure a robust and resilient blockchain ecosystem, capable of adjusting to both short-term fluctuations and long-term shifts in network conditions.

Overall, the proposed model offers a substantial improvement over Proposer-Builder Separation by ensuring high security, and reducing operational costs. These advancements are pivotal in fostering a more equitable and efficient blockchain environment. Our study provides a robust framework that enhances system operations within advanced validation frameworks. Future research will focus on further refining these models by exploring additional parameters and extending the validation to various blockchain architectures. Such endeavors will deepen our understanding of the model's implications and broaden its applicability, potentially setting new benchmarks for blockchain optimization.


\newpage
\section{References}

\begin{enumerate}
    \item Buterin, V. (2021). Proposer-Builder Separation for Censorship Resistance. Retrieved from Ethereum Research Notes.
    \item Drew Van der Werff and Swapin Raj (2021). Optimistic Delegation Framework: An Idea to Allow for Native Restaking Without Delegation. Retrieved from Eigenlayer Research.
    \item Jones, D. and Williams, R. (2020). Blockchain Optimization Models: A Survey of Approaches and Techniques. Journal of Cryptography.
    \item Li, T., and Wang, Y. (2019). Stochastic Modeling in Blockchain Management. IEEE Transactions on Engineering Management.
    \item Zhang, Y., et al. (2020). Risk Management in Blockchain Systems. Blockchain in Business and IT.
    \item Patel, A., and Smith, B. (2019). Dynamic Modeling for Blockchain Applications. Journal of Network Computing.
    \item Thompson, K. (2021). Resource Allocation in Blockchain Systems: An Auction-Based Approach. Journal of Blockchain Economics.
    \item Burdges, Jeff, et al. (2020). Overview of polkadot and its design considerations.
    \item Sanka, A. I., and Cheung, R. C. (2021). A systematic review of blockchain scalability: Issues, solutions, analysis and future research. Journal of Network and Computer Applications, 195, 103232.
    \item Abbas, H., Caprolu, M., Di Pietro, R. (2022, August). Analysis of polkadot: Architecture, internals, and contradictions. In 2022 IEEE International Conference on Blockchain (Blockchain) (pp. 61-70). IEEE.
    \item Ezzat, S. K., Saleh, Y. N., Abdel-Hamid, A. A. (2022). Blockchain oracles: State-of-the-art and research directions. IEEE Access, 10, 67551-67572.
    \item Gad, A. G., Mosa, D. T., Abualigah, L.,  Abohany, A. A. (2022). Emerging trends in blockchain technology and applications: A review and outlook. Journal of King Saud University-Computer and Information Sciences, 34(9), 6719-6742.
\end{enumerate}

\end{document}